\documentclass[10pt,letterpaper]{article}
\usepackage{opex3}
\usepackage{SIunits}
\usepackage{epstopdf}

\begin{document}

\title{Frequency converter implementing\\
       an optical analogue of\\
       the cosmological redshift}

\author{Vincent Ginis,$^1$ Philippe Tassin,$^{1,2}$ Ben Craps,$^3$\\ and Irina Veretennicoff$^{1,3}$}
\address{$^1$Department of Applied Physics and Photonics, Vrije Universiteit Brussel,\\
             Pleinlaan 2, B-1050 Brussel, Belgium}
\address{$^2$Ames Laboratory-U.S. DOE, and Department of Physics and Astronomy,\\
             Iowa State University, Ames, Iowa 50011, USA}
\address{$^3$Theoretische Natuurkunde, Vrije Universiteit Brussel, and\\ The International Solvay Institutes,
             Pleinlaan 2, B-1050 Brussel, Belgium}

\email{vincent.ginis@vub.ac.be}

\begin{abstract} 
According to general relativity, the frequency of electromagnetic radiation is altered by the expansion of the universe. This effect---commonly referred to as the cosmological redshift---is of utmost importance for observations in cosmology. Here we show that this redshift can be reproduced on a much smaller scale using an optical analogue inside a dielectric metamaterial with time-dependent material parameters. To this aim, we apply the framework of transformation optics to the Robertson-Walker metric. We demonstrate theoretically how perfect redshifting or blueshifting of an electromagnetic wave can be achieved without the creation of sidebands with a device of finite length.
\end{abstract}

\ocis{(160.3918) Metamaterials; (260.2110) Electromagnetic optics; (060.2630) Frequency modulation; (50.5720) Relativity.}


\section{Introduction}
Recently, there has been considerable interest in developing optical analogues of black holes in table-top devices~\cite{Philbin-2008, Zhang-2009,Narimanov-2009,Cui-2009}. The electromagnetic parameters of these devices are designed in such a way as to mimic effects arising in the presence of black holes. Using this approach, an artificial event horizon has been implemented by means of a moving medium~\cite{Philbin-2008}, an inhomogeneous refractive index profile was proposed to mimic celestial mechanics~\cite{Zhang-2009}, and a broadband omnidirectional absorber has been developed~\cite{Narimanov-2009,Cui-2009}. Although the equivalence between gravitational effects and optical media was suggested by several authors~\cite{Balazs-1957,Plebanski-1960,Felice-1971}, it has recently been conceived from a different point of view~\cite{Ward-1996} and has become a tool to design optical devices within the framework of transformation optics~\cite{Pendry-2006,Leonhardt-2006,Leonhardt-2006-2, Leonhardt-2009}.

Transformation optics provides a new way of looking at the interaction between light and matter. This recently developed method in electromagnetism is based on the analogy between the macroscopic Maxwell's equations in complex materials and the free-space Maxwell's equations on the background of an arbitrary metric. It describes how a general topological deformation of reality can be implemented using materials with nontrivial constitutive parameters. The invisibility cloak, which bends the electromagnetic reality, i.e., the electromagnetic energy flow, around a hole---which therefore turns invisible---, is probably the most famous example of this technique~\cite{Pendry-2006, Leonhardt-2006,Cai-2007,Valentine-2009}. Moreover, the cloaking principle extends beyond electromagnetism and can be applied in acoustics~\cite{Cummer-2008}, hydrodynamics~\cite{Farhat-2008}, and even quantum mechanics~\cite{Zhang-2008}.

Up to now, most applications of transformation optics have been related to the spatial manipulation of optical beams and  the design of imaging systems \cite{Rahm2-2008, Kwon-2008, Jacob-2006}. These devices are generated through spatial coordinate transformations and do not consider curved geometries. Transformation optics, however, applies to general four-dimensional metrics, and thus extends beyond the scope of existing applications. In this paper, we show how a four-dimensional metric that plays a role in general relativity---the Robertson-Walker metric---can be implemented directly with dielectrics and we demonstrate theoretically that the cosmological redshift can be reproduced inside a dielectric structure of finite size with time-varying electromagnetic properties.

\section{Electromagnetic analogue of the cosmological redshift}
The cosmological redshift of an electromagnetic wave traveling from one place to another is an effect that originates from the expansion of the universe. Such an expanding background can be modeled by the Robertson-Walker metric,
\begin{equation}
ds^2 = -c^2\mathrm{d}t^2 + a^2(t)\left[\frac{\mathrm{d}r^2}{1-\kappa r^2} + r^2\, d\Omega^2\right],
\end{equation}
where $a(t)$ is a real, dimensionless scale factor, related to the well-known Hubble parameter by $H = \dot{a}/a$, and $\kappa$ represents the spatial curvature of the metric. An electromagnetic wave with frequency $\omega_\mathrm{em}$ emitted at instance $t = t_\mathrm{em}$ will be observed at a different frequency $\omega_\mathrm{obs}$ at time $t = t_\mathrm{obs}$. These two frequencies are related by (see, e.g., Ref.~\cite{Carroll-2003})
\begin{equation}
\frac{\omega_\mathrm{obs}}{\omega_\mathrm{em}} = \frac{a(t_\mathrm{em})}{a(t_\mathrm{obs})}.
\label{Eq:CosmologicalRedshift}
\end{equation}
In an expanding universe, the frequency of electromagnetic waves is redshifted.

\begin{figure}
  \begin{center}
  \includegraphics[height=3.1cm]{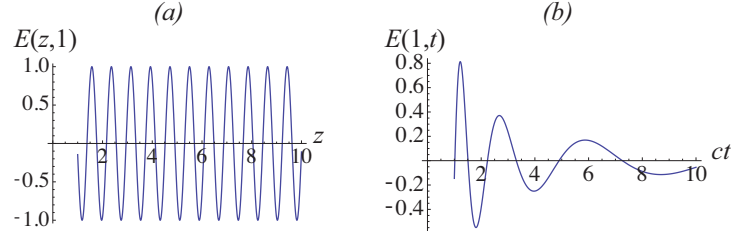}\label{Fig:RobertsonWalkerSolutionZ}
\end{center}
\caption{A graphical representation of a traveling wave solution inside a medium that is the electromagnetic analogue of the Robertson-Walker metric. (a) The spatial variation at constant time $t=1$ and (b) the temporal variation of this solution for fixed location $z = 1$, when we modulate the permittivity and permeability with $a(t) = 1 + t$.}
  \label{Fig:RobertsonWalkerSolution}
\end{figure}

To construct an electromagnetic analogue of this effect, we proceed by calculating the material properties of a dielectric medium using transformation optics~\cite{Pendry-2006,Leonhardt-2006-2, Leonhardt-2009}. It is straightforward to show that the Robertson-Walker metric without spatial curvature ($\kappa=0$) can be translated into an isotropic, homogeneous dielectric with constitutive parameters
\begin{equation}
\epsilon^x_{\phantom{x}x} = \mu^x_{\phantom{x}x} = a(t),\;\;
\epsilon^y_{\phantom{y}y} = \mu^y_{\phantom{y}y} = a(t),\;\;
\epsilon^z_{\phantom{z}z} = \mu^z_{\phantom{z}z} = a(t).\label{Eq:RobertsonWalkerMaterials}
\end{equation}
The traveling wave solutions inside such a medium are found by solving Maxwell's equations in the absence of free sources and currents.
In combination with Eq.~(\ref{Eq:RobertsonWalkerMaterials}), Maxwell's equations combine into a generalized wave equation,
\begin{equation}
\Delta\mathbf{E} - \frac{1}{c^2}\frac{\partial}{\partial t}\left(a(t)\frac{\partial}{\partial t}(a(t) \mathbf{E})\right) = 0,
\end{equation}
where $c$ is the speed of light in vacuum. Its solutions are waves with time-dependent amplitude propagating along an arbitrary direction labeled by the unit vector $\mathbf{1}_\mathrm{k}$:
\begin{equation}
\label{Eq:Superposition}
\mathbf{E}(\mathbf{r},t) = \frac{1}{a(t)}\, G\left(\mathbf{r}\cdot\mathbf{1}_\mathrm{k}- c \int_{t_0}^t \frac{\mathrm{d}\tilde{t}}{a(\tilde{t})}\right)\mathbf{1_\mathrm{E}},
\end{equation}
where $G(x)$ is an arbitrary differentiable function. To gain some insight in these traveling waves, we consider a harmonic function $G(x)=\cos(x)$ in Fig.~\ref{Fig:RobertsonWalkerSolution} and we plot the temporal and spatial variation of Eq.~(\ref{Eq:Superposition}). We see that the wave has a well-defined wavelength, but that its frequency is changing in time. A straightforward calculation of the instantaneous frequency $\omega_\mathrm{inst}(t) = -\partial_t(\mathbf{k}\cdot\mathbf{r} - |\mathbf{k}| c \int_{t_0}^t \mathrm{d}\tilde{t}/a(\tilde{t}))$ reveals that the frequencies at two different times $t_1$ and $t_2$ satisfy $\omega_\mathrm{inst}(t_2)/\omega_\mathrm{inst}(t_1) = a(t_1)/a(t_2)$, in agreement with the redshift formula of Eq.~(\ref{Eq:CosmologicalRedshift}).


\section{A frequency converter of finite size}

\begin{figure}
  \begin{center}
    \includegraphics[height=5.5cm]{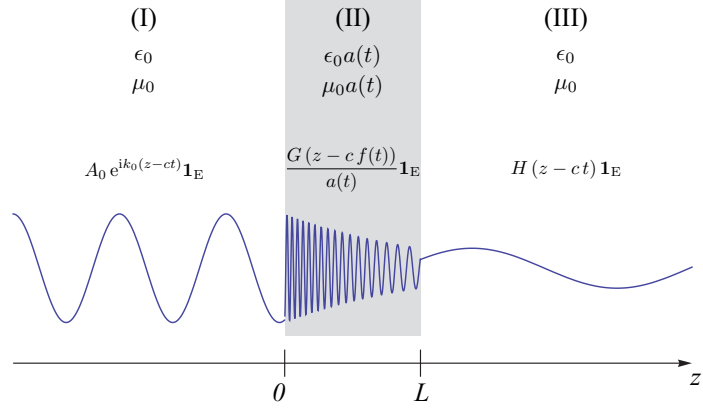}
  \end{center}
  \caption{A setup where we implement a finite Robertson-Walker device. At the leftmost boundary a monochromatic wave of frequency $\omega_0$ and amplitude $A_0$ illuminates the device. We modulate the permittivity and permeability as $\epsilon(t) = \mu(t) = a(t) = 1 + t$. We show that the output undergoes a frequency shift according to the cosmological redshift formula.}
  \label{Fig:RobertsonWalkerSetup}
\end{figure}

As shown above, a traveling wave in an infinite Robertson-Walker medium is subject to an equivalent of the cosmological redshift. This does not guarantee, however, that we can use this medium to alter the frequency of an incident wave, since any practical application requires materials of finite extent. Furthermore, as was pointed out in Ref.~\cite{Rahm-2008}, the boundary conditions play a crucial role in a transformation-optical device of finite extent. We therefore examine in this section whether a slab of the Robertson-Walker medium with thickness $L$ (see Fig.~\ref{Fig:RobertsonWalkerSetup}) would suit as a frequency converter. To simplify notation, we introduce the function
\begin{equation}
\label{Eq:FrequencyTunerFt}
f(t) = \int_{t_0}^t\frac{\mathrm{d}\tilde{t}}{a(\tilde{t})}.
\end{equation}

\begin{figure}
  \begin{center}
	\includegraphics[height=5.0cm]{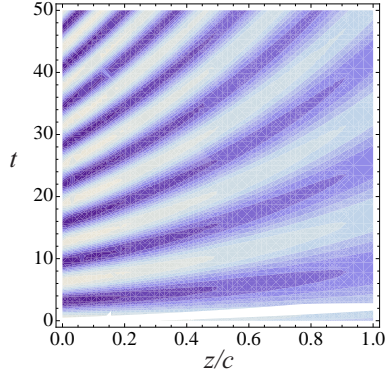}
  \end{center}
  \caption{The wave fronts inside the frequency converter as a function of space and time when the leftmost boundary is illuminated by a monochromatic wave. In this diagram, we have simulated a linear evolution of the scale factor $a(t) = t$.}
  \label{Fig:ztDiagram}
\end{figure}

We illuminate the dielectric from the left in region $(\mathrm{I})$ with a monochromatic plane wave of frequency $\omega_0 = |\mathbf{k}_0| c$, and we want to calculate the wave that is emitted in region $(\mathrm{III})$.
In general, each interface would give rise to transmitted and reflected waves, and one would explicitly impose 
continuity of the tangential components of both $\mathbf{E}$ and $\mathbf{H}$. However, since the impedance $\eta = (\mu/\epsilon)^{1/2}$ is the same on both sides of each interface, we can anticipate that there are no 
reflected waves. We therefore restrict our attention to the electric field $\mathbf{E}$, consider purely right-moving waves and impose continuity at each interface. In regions $(\mathrm{I})$, $(\mathrm{II})$, and $(\mathrm{III})$, respectively, the electric field can be written as
\begin{eqnarray}
\label{Eq:FieldInRegion2}
\mathbf{E}_\mathrm{I}(z,t) &=& A_0\,\mathrm{e}^{\mathrm{i}k_0 \left(z-ct\right)}\mathbf{1_\mathrm{E}},\\
\mathbf{E}_\mathrm{II}(z,t) &=& \frac{G\left(z-cf(t)\right)}{a(t)}\mathbf{1_\mathrm{E}},\\
\label{Eq:FieldInRegion3}
\mathbf{E}_\mathrm{III}(z,t) &=& H\left(z-c\, t\right)\mathbf{1_\mathrm{E}},
\end{eqnarray}
where $G(x)$ and $H(x)$ are differentiable functions. The continuity of the tangential component of the electric field at $z = 0$ yields
\begin{equation}
A_0\mathrm{e}^{-\mathrm{i} k_0 c t} = \frac{G\left(-cf(t)\right)}{a(t)}.
\end{equation}
This equation can be used to evaluate $G(x)$ for an arbitrary value of its argument $x$:
\begin{equation}
\label{Eq:WaveInsideComponent}
G(x) = a\left(f^{-1}\left(-\frac{x}{c}\right)\right) A_0\mathrm{e}^{-\mathrm{i} k_0 c f^{-1}\left(-\frac{x}{c}\right)},
\end{equation}
where we have introduced the inverse function $f^{-1}$. In Fig.~\ref{Fig:ztDiagram}, we plot the evolution of the wavefronts inside the material as a function of space and time. We can now combine Eq.~(\ref{Eq:FieldInRegion2}) with Eq.~(\ref{Eq:WaveInsideComponent}) to determine the electric field at the rightmost boundary ($z=L$) of the device:
\begin{equation}
\mathbf{E_\mathrm{II}}(L,t) =  \frac{a\left(f^{-1}\left(f(t)-\frac{L}{c}\right)\right)}{a(t)}A_0\mathrm{e}^{-\mathrm{i} k_0 c f^{-1}\left(f(t)-\frac{L}{c}\right)}\mathbf{1_\mathrm{E}}.
\end{equation}
To retrieve a general expression for the electric field in region ($\mathrm{III}$), we have to express the continuity of the electric field at the rightmost boundary and we ultimately find that
\begin{equation}
\mathbf{E_\mathrm{III}}(z,t) = \mathbf{E_\mathrm{II}}\left(L,t-\frac{z-L}{c}\right).
\end{equation}

The instantaneous frequency of the electromagnetic wave emitted in the vacuum region $(\mathrm{III})$ behind the device ($z=L$) can be written as
\begin{equation}
\label{Eq:FrequencyShift}
\omega_\mathrm{out}  = \omega_0 \frac{a\left(f^{-1}\left(f(t)-\frac{L}{c}\right)\right)}{a(t)}.
\end{equation}
This equation shows that the proposed design from Fig.~\ref{Fig:RobertsonWalkerSetup} indeed executes a frequency shift. Since the velocity of the wavefronts $v = c/\sqrt{\epsilon\mu}=c/a(t)$, we know that $L = \int_{t_1}^{t_2} c/a(t)\, \mathrm{d}t = c\left(f(t_2)-f(t_1)\right)$, where $t_1$ and $t_2$ indicate the time of incidence and departure of a wavefront. If we are observing at time $t$ at the rightmost boundary, then $f^{-1}\left(f(t)-L/c\right)$ corresponds to the time when the wavefront was at the leftmost boundary. Hence, we retrieve the cosmological redshift formula, ${\omega_\mathrm{inst}(t_2)}/{\omega_\mathrm{inst}(t_1)} = {a(t_1)}/{a(t_2)}$. The reader should also note that the amplitude is altered with a similar factor, so that the device might also be used as an optical amplifier. It will often be beneficial to create a constant frequency shift, i.e., to generate a monochromatic wave in region (III); in this case, the right hand side of Eq.~(\ref{Eq:FrequencyShift}) has to be constant. 
By renaming $f(t)= x$, this condition immediately implies that $a(f^{-1}(x)) = \exp(\alpha x) p(x)$, where $\alpha$ is an arbitrary (real) constant and $p(x)$ is a periodic function with period $L/c$. 
Using Eq. (\ref{Eq:FrequencyTunerFt}), we observe that $a(f^{-1}(x))=(f^{-1})^\prime (x)$ and infer that the condition is satisfied when $f^{-1}(x)=\exp(\alpha x) q(x) - \beta$ with $\beta$ constant and $q(x)$ periodic with period $L/c$. By inverting 
this relation to obtain $f(t)$ and differentiating with respect to $t$, we can calculate those $a(t)$ which generate a monochromatic wave in region (III). For instance, the special case $q(x)=1$ leads to $f(t)=\ln(t+\beta)/\alpha$ and thus to the linear profile $a(t)=\alpha(t+\beta)$.
The resulting frequency shift is then given by
\begin{equation}
\frac{\omega_\mathrm{out}}{\omega_0} = \mathrm{e}^{-\alpha L/c}.
\label{Eq:DeviceFormula}
\end{equation}

\section{Conclusion and discussion}
We note that the output signal has all its energy at the desired frequency without the creation of sidebands. The conversion is thus characterized by a very high efficiency. The proposed device achieves a frequency shift using \emph{linear} materials, so that the superposition principle remains valid and an arbitrary wavepacket can be frequency-shifted. This frequency shift is possible due to the time evolution of its parameters, which renders it non-stationary \cite{Budko-2009}.

Equation~(\ref{Eq:DeviceFormula}) enables us to calculate the characteristics of a device that generates the desired frequency shift of a monochromatic wave. Let us suppose, for instance, that we want to shift an incident frequency of $f_0 = \unit{100}{\tera\hertz}$ by \unit{10}{\giga\hertz} in a material of thickness $L = 2 \lambda_0 = \unit{6}{\micro\meter}$. We then find that the modulation rate of the refractive index must equal $\alpha \approx \unit{5}{\giga\hertz}$. Since the corresponding wavelength of this modulating signal is much larger than the dimensions of the device, we can generate the permittivity variation by low-frequency electro-optic modulation. The most difficult part is the time evolution of the permeability. This can be achieved by introducing an array of split-ring resonators \cite{Smith-2004, Soukoulis-2006} in the electro-optic material. 
The frequency-dependent permeability of an array of split rings can be modeled by
\begin{equation}
\mu(\omega)=1+\frac{F\omega^2}{\omega_\mathrm{LC}^2-\omega^2},
\end{equation}
where $F$ is the filling factor and $\omega_\mathrm{LC}$ is the resonance frequency, which is inversely proportional to the square root of the material's permittivity $\epsilon_\mathrm{C}$ inside the gap of the split rings. Operating far below resonance, we can approximate the permeability by 
$\mu(\omega)\approx1+F\omega^2 c^2 l^2 \epsilon_\mathrm{C}/(dw),$
with $l$, $d$, $w$ the geometrical parameters of the split rings \cite{Soukoulis-2006}. By properly choosing the electro-optic material inside the gap of the split rings, we can thus vary the permeability and permittivity according to Eq.~(\ref{Eq:RobertsonWalkerMaterials}) with electro-optics to establish a frequency shift determined by Eq.~(\ref{Eq:DeviceFormula}).

\section*{Acknowledgments} 
Work at the VUB was supported by the Belgian Science Policy Office (Grant No. IAP VI/10 Photonics@be \& Grant No. IAP VI/11), by the FWO-Vlaanderen (Project G011410N, fellowships), and by the Research Council (OZR) of the VUB. Work at Ames Laboratory was supported by the Department of Energy (Basic Energy Sciences) under Contract No. DE-AC02-07CH11358. P.~T. acknowledges the Belgian American Educational Foundation for financial support.


\begin{thebibliography}{10}
\newcommand{\enquote}[1]{``#1''}

\bibitem{Philbin-2008}
T.~G. Philbin, C.~Kuklewicz, S.~Robertson, S.~Hill, F.~Konig, and U.~Leonhardt,
  \enquote{{Fiber-Optical Analog of the Event Horizon},} Science \textbf{319},
  1367--1370 (2008).

\bibitem{Zhang-2009}
D.~A. {Genov}, S.~Zhang, and X.~Zhang, \enquote{Mimicking celestial mechanics
  in metamaterials,} Nature Phys. \textbf{5}, 687--692 (2009).

\bibitem{Narimanov-2009}
E.~E. Narimanov and A.~V. Kildishev, \enquote{Optical black hole: Broadband
  omnidirectional light absorber,} Appl. Phys. Lett. \textbf{95}, 041106
  (2009).

\bibitem{Cui-2009}
C.~Qiang and C.~T. Jun, \enquote{An electromagnetic black hole made of
  metamaterials,} arXiv:0910.2159v1 [physics.optics]  (2009).

\bibitem{Balazs-1957}
N.~L. {Balazs}, \enquote{Effect of a gravitational field, due to a rotating
  body, on the plane of polarization of an electromagnetic wave,} Phys.\ Rev.
  \textbf{110}, 236--239 (1957).

\bibitem{Plebanski-1960}
J.~{Plebanski}, \enquote{Electromagnetic waves in gravitational fields,} Phys.\
  Rev. \textbf{118}, 1396--1408 (1960).

\bibitem{Felice-1971}
D.~F. {Felice}, \enquote{On the gravitational field acting as an optical
  medium,} Gen.\ Rel.\ Grav. \textbf{2}, 347--357 (1971).

\bibitem{Ward-1996}
A.~J. {Ward} and J.~B. {Pendry}, \enquote{Refraction and geometry in
  {M}axwell's equations,} J.\ Mod.\ Phys. \textbf{43}, 773--793 (1996).

\bibitem{Pendry-2006}
J.~B. {Pendry}, D.~{Schurig}, and D.~R. {Smith}, \enquote{Controlling
  electromagnetic fields,} Science \textbf{312}, 1780--1782 (2006).

\bibitem{Leonhardt-2006}
U.~{Leonhardt}, \enquote{Optical conformal mapping,} Science \textbf{312},
  1777--1780 (2006).

\bibitem{Leonhardt-2006-2}
U.~{Leonhardt} and T.~G. {Philbin}, \enquote{General relativity in electrical
  engineering,} New\ J.\ Phys. \textbf{8}, 247--264 (2006).

\bibitem{Leonhardt-2009}
U.~{Leonhardt} and T.~G. {Philbin}, \enquote{Transformation optics and the
  geometry of light,} Prog.\ Opt. \textbf{53}, 70--152 (2009).

\bibitem{Cai-2007}
W.~{Cai}, U.~K. {Chettiar}, A.~V. {Kildishev}, and V.~M. {Shalaev},
  \enquote{Optical cloaking with metamaterials,} Nature Photon. \textbf{1},
  224--227 (2007).

\bibitem{Valentine-2009}
J.~{Valentine}, J.~{Li}, T.~{Zentgraf}, G.~{Bartal}, and X.~{Zhang},
  \enquote{An optical cloak made of dielectrics,} Nature Mater. \textbf{8},
  568--571 (2009).

\bibitem{Cummer-2008}
S.~A. {Cummer}, B.-I. {Popa}, D.~{Schurig}, D.~R. {Smith}, J.~{Pendry},
  M.~{Rahm}, and A.~{Starr}, \enquote{Scattering theory derivation of a 3{D}
  acoustic cloaking shell,} Phys.\ Rev.\ Lett. \textbf{100}, 024301 (2008).

\bibitem{Farhat-2008}
M.~{Farhat}, S.~{Enoch}, S.~{Guenneau}, and A.~B. {Movchan}, \enquote{Broadband
  cylindrical acoustic cloak for linear surface waves in a fluid,} Phys.\ Rev.\
  Lett. \textbf{101}, 134501 (2008).

\bibitem{Zhang-2008}
S.~{Zhang}, D.~A. {Genov}, C.~{Sun}, and X.~{Zhang}, \enquote{Cloaking of
  matter waves,} Phys.\ Rev.\ Lett. \textbf{100}, 123002 (2008).

\bibitem{Rahm2-2008}
M.~{Rahm}, D.~A. {Roberts}, J.~B. {Pendry}, and D.~R. {Smith},
  \enquote{Transformation-optical design of adaptive beam bends and beam
  expanders,} Opt.~Express \textbf{16}, 11555--11567 (2008).

\bibitem{Kwon-2008}
D.~{Kwon} and D.~H. {Werner}, \enquote{Polarization splitter and polarization
  rotator designs based on transformation optics,} Opt.~Express \textbf{16},
  18731--18738 (2008).

\bibitem{Jacob-2006}
Z.~{Jacob}, L.~V. {Alekseyev}, and E.~{Narimanov}, \enquote{Optical hyperlens:
  Far-field imaging beyond the diffraction limit,} Opt.~Express \textbf{14},
  8247--8256 (2008).

\bibitem{Carroll-2003}
S.~{Carroll}, \emph{Spacetime and Geometry} (Addison Wesley, New York, 2003).

\bibitem{Rahm-2008}
M.~{Rahm}, D.~{Schurig}, D.~A. {Roberts}, S.~A. {Cummer}, D.~R. {Smith}, and
  J.~B. {Pendry}, \enquote{Design of electromagnetic cloaks and concentrators
  using form-invariant coordinate transformations of {M}axwell's equations,}
  Photon.~Nanostruct.: Fundam. Applic. \textbf{6}, 87--95 (2008).

\bibitem{Budko-2009}
N.~V. {Budko}, \enquote{Electromagnetic radiation in a time-varying background
  medium,} Phys. Rev. A \textbf{80}, 053817 ({2009}).

\bibitem{Smith-2004}
D.~R. Smith, J.~B. Pendry, and M.~C.~K. Wiltshire, \enquote{{Metamaterials and
  Negative Refractive Index},} Science \textbf{305}, 788--792 (2004).

\bibitem{Soukoulis-2006}
C.~M. Soukoulis, M.~Kafesaki, and E.~N. Economou, \enquote{Negative-index
  materials: New frontiers in optics,} Adv. Mater. \textbf{18}, 1941--1952
  (2005).

\end{thebibliography}
\end{document}